\documentclass[AMA,STIX1COL]{WileyNJD-v2}

\articletype{Article Type}%

\received{26 April 2016}
\revised{6 June 2016}
\accepted{6 June 2016}

\usepackage{IEEEtrantools}
\usepackage{bm}
\usepackage{amsmath}

\usepackage{pifont}
\usepackage{makecell}
\newcommand{\xmark}{\ding{55}}%

\usepackage{subcaption}

\newcommand{\mbeq}{\overset{!}{=}}
\begin{document}

\title{Principled Drug-Drug Interaction Terms for Bayesian Logistic Regression Models of Drug Safety in Oncology Phase I Combination Trials}

\author[1]{Lukas A. Widmer*}

\author[2]{Andrew Bean} 

\author[2]{David Ohlssen}

\author[1]{Sebastian Weber}

\authormark{WIDMER \textsc{et al}} %

\address[1]{\orgname{Novartis Pharma AG}, \orgaddress{CH-4002 \state{Basel}, \country{Switzerland}}} 

\address[2]{\orgdiv{Novartis Pharmaceuticals}, \orgaddress{East Hanover, \state{NJ}, \country{USA}}}

\corres{*~Lukas A. Widmer, Novartis Pharma AG, CH-4002 Basel, Switzerland. \email{lukas\_andreas.widmer@novartis.com}}


\abstract[Summary]{
In Oncology, trials evaluating drug combinations are becoming more common. While combination therapies bring the potential for greater efficacy, they also create unique challenges for ensuring drug safety. In Phase-I dose escalation trials of drug combinations, model-based approaches enable efficient use of information gathered, but the models need to account for trial complexities: appropriate modeling of interactions becomes increasingly important with growing numbers of drugs being tested simultaneously in a given trial. In principle, we can use data from multiple arms testing varying combinations (like a backbone treatment combined with different candidate combination partners) to jointly estimate toxicity of the drug combinations. However, such efforts have highlighted limitations when modelling drug-drug interactions in the Bayesian Logistic Regression Model (BLRM) framework used to ensure patient safety. Previous models either do not account for non-monotonicity due to antagonistic toxicity, or exhibit the fundamental flaw of exponentially overpowering the contributions of the individual drugs in the dose-response. This specifically leads to issues when drug combinations exhibit antagonistic toxicity, in which case the toxicity probability gets vanishingly small as doses get very large.

We put forward additional constraints inspired by Paracelsus' intuition of "the dose makes the poison" which avoid this flaw and present an improved interaction model which is compatible with these constraints. We create instructive data scenarios that showcase the improved behavior of this more constrained drug-drug interaction model in terms of preventing further dosing at overly toxic dose combinations and more sensible dose-finding under antagonistic drug toxicity. This model is now available in the open-source OncoBayes2 R package that implements the BLRM framework for an arbitrary number of drugs and trial arms. 
}


\jnlcitation{\cname{%
\author{L. A. Widmer},
\author{A. Bean},
\author{D. Ohlssen},
and
\author{S. Weber}} (\cyear{2023}), 
\ctitle{Principled Drug-Drug Interaction Terms for Bayesian Logistic Regression Models of Drug Safety in Oncology Phase I Combination Trials}, 
\cjournal{TBD}, 
\cvol{2023;TBD}.
}

\maketitle


\newpage
\section{Introduction}\label{sec1}
In Phase-I Oncology, dose escalation trials are concerned with striking the right balance of ensuring patient safety while rapidly escalating drug doses to avoid exposure of patients to sub-therapeutic dosing and determining the maximum tolerated dose\cite{LeTourneau2009}. This is motivated by the cytotoxic nature of (older) cancer drugs, where both efficacy and toxicity were expected to increase with dose. In contrast to drugs targeting other diseases, lack of efficacy leads to death, and as a consequence, some level of toxicity is acceptable. 
The goals of such dose-escalation studies are to systematically increase dosing
\begin{enumerate}
\item as quickly as possible, such that a biologically active (and potentially efficacious) dose is reached as soon as possible, and to keep overall study size small,
\item to determine the recommended phase II dose (RP2D), for instance, via the maximum tolerated dose (MTD), for use in future trials\cite{Neuenschwander2014},
\item while doing so as safely as possible, i.e., while controlling the rate of dose-limiting toxicity (DLT) events during the trial, as well as controlling the probability of declaring an overly toxic final dose (which would be a risk to any downstream trials).
\end{enumerate}
Typically, such escalation is performed in treatment cycles (e.g. of length 4 weeks), where patients are treated according to a defined schedule. DLT events are recorded for the duration of the treatment cycle, and based on these observations a dose for the next cohort of patients is chosen. Combination therapies have become a quite common strategy to (among other considerations) increase efficacy \cite{Mokhtari2017, Palmer2017} while maintaining an acceptable safety profile. Thus, we may also want to explore the safety of the compound of interest in combination with other compounds (such as an established standard of care), necessitating modelling safety-relevant drug-drug interactions on top of the single drug-toxicity relationships.

A fit-for-purpose approach for modelling and running such combination therapy trials is adaptive escalation with overdose control (EWOC) informed by a Bayesian logistic regression model (BLRM)\cite{Neuenschwander2014}. Initially, we typically have limited knowledge about toxicity for a given dose or dose combination (e.g. from single-drug data and pre-clinical drug-drug interaction data), so only small cohorts may be recruited to limit the risk to participants \cite{Babb1998}. After enrolling a few patients (typically three to six) into a chosen dose level, we observe how many patients experience DLT events during one treatment cycle. At this stage we update the model with the newly-acquired data and decide what to do next: either escalate, retain or de-escalate the dose, or stop the trial. 

In modern Oncology, we apply model-based approaches to more efficiently conduct trials. Both for single-drug and combinations, these approaches model the dose-toxicity relationship in order to limit the probability of administering an overdose, for instance using the EWOC criterion. Since combination therapies have become rather common, the ability of any model-based approach to properly account for drug-drug interactions of those drug combinations has become increasingly important. Two drugs can exhibit synergistic toxicity, i.e., combining them leads to higher toxicity than expected from the two drugs individually, or antagonistic toxicity, i.e., combining them leads to lower toxicity than expected from the two drugs individually (we will formalize this later on). In the case of the BLRM, two and three-drug interaction models have been developed \cite{Neuenschwander2014} and more recently been extended to $N$ drugs \cite{OncoBayes2}. Other popular single-drug safety models, such as the Continual Reassessment Method (CRM)\cite{Quigley1990} or Bayesian optimal interval design (BOIN)\cite{Yuan2016}, also have more recent extensions for modelling drug combinations (e.g., the POCRM\cite{Wages2011} or BOIN-COMB\cite{Zhang2016, Lin2017}).

Here, we will detail a principled drug-drug interaction model for the BLRM method. This model is now implemented as the new default drug-drug interaction model (from version 0.8-0 onwards) in the open-source OncoBayes2\cite{OncoBayes2} R package available on CRAN. Note that the focus of this work is on principles and illustrative examples using data scenarios, and explicitly \emph{not} to compare the principled model against others in operating characteristics simulation studies, since the outcome of such simulations depends heavily on the design of the particular trial, scenarios chosen, and comparing methods that are applicable. We provide code to run the models on the data scenarios detailed in this manuscript as well as produce the figures using the OncoBayes2 R package, which can be re-used to implement custom scenarios or simulations relevant to the trial design of interest. Thus, our focus is on deriving an improved drug-drug interaction model by formalizing desirable properties and designing a drug-drug interaction model for the BLRM that satisfies them. In principle, these properties can also be used as guidance to design drug-drug interaction models for other methods.

The remainder of this article is structured as follows. Section 2 describes the BLRM methodology for single drugs, multiple drugs without and finally with drug-drug interactions. Section 3 shows the simulated case studies that show the advantages and limitations of the proposed approach. The article concludes with a discussion in Section 4.

\section{Methodology}\label{sec:methodology}
As in previous work \cite{Neuenschwander2014}, our data consists of cohorts of $n_{d}$ patients at dose ${d}$. Note that typically, we pre-specify the dose levels we want to test each drug at as in related literature, however, we omit the index for the dose level here and for simplicity assume a continuous dose $d$. We then observe $r_{d}$ patients exhibiting a DLT event with a probability of $\pi(d)$ over the course of their first treatment cycle. We observe binomially-distributed cohorts, i.e.,
\begin{IEEEeqnarray}{rCl}
r_{d} | \pi(d), n_{d} & \sim & \text{Binomial}\left(\pi(d), n_{d}\right).
\end{IEEEeqnarray}
Typically, three intervals of $\pi(d)$ are considered\cite{Neuenschwander2014}:
\begin{IEEEeqnarray*}{lrCl}
\text{Underdosing:}\quad & \pi(d) & < & 16 \% \\
\text{Target toxicity:}\quad & 16 \% \leq & \pi(d) & < 33 \%  \\
\text{Overdosing:}\quad & 33 \% & \leq & \pi(d)
\end{IEEEeqnarray*}
In the underdosing category, we would be willing to accept slightly higher toxicity in trade for higher efficacy, given the severity of the disease. The target toxicity then is typically between 16 \% and 33 \%, and DLT probabilities higher than 33 \% are considered overdosing, i.e., unacceptably toxic. The EWOC criterion\cite{Babb1998} then puts an upper bound \eqref{eq:EWOC} on the probability of choosing an overdose, typically to 25\% or below:
\begin{IEEEeqnarray}{rCl}
\text{EWOC fulfilled} \Longleftrightarrow P(\pi(d) \geq 33\%) & \leq & 25\% \label{eq:EWOC}. 
\end{IEEEeqnarray}

\subsection{Single-drug model}
To model the dose-DLT-probability response $\pi(d) = P(\text{DLT from drug} \text{ at (continuous) dose } d)$, we use the standard logistic regression model in terms of the normalized log-dose \cite{Babb1998, Neuenschwander2014}, and model $\pi$ on the log-odds scale as
\begin{IEEEeqnarray}{rCl}
\text{logit}(\pi(d)) & = & \text{log}(\alpha) + \beta \, \text{log}\left(\frac{d}{d^*}\right).
\end{IEEEeqnarray}
Here, $d^*$ is the reference dose for the drug which is used to normalize the dose $d$, and the parameters $\alpha$ and $\beta$ have the following interpretation:
\begin{itemize}
\item $\alpha \geq 0$ corresponds to the odds of observing a DLT event at the reference dose $d^*$: 
\begin{equation*}
	\alpha  = \pi(d^*)/(1-\pi(d^*)).
\end{equation*}
\item $\beta > 0$ is the slope of the logistic curve, and determines how quickly toxicity increases with increasing dose. More formally, if we increase the dose $d \neq 0$ by a factor of $\gamma$, then the odds of observing a DLT event will be increased by a factor of $\gamma^\beta$.
\end{itemize}
Notable properties of this model include monotonicity ($\pi(d)$ is non-decreasing) and appropriate limiting behavior for small and large doses ($\lim_{d\to 0} \pi(d) = 0$ and $\lim_{d\to\infty} \pi(d) = 1$).

\subsection{Multi-drug model without interactions}
The single-drug model can be extended to multiple drugs, under the assumption that all drugs generate DLT events independently (i.e., we assume no drug-drug interactions\cite{Neuenschwander2014}), giving rise to the probability $\pi_\perp$. The simplest case is the one for two drugs, e.g., 1 and 2, that generate DLT events independently. The overall probability of observing no toxicity from drug 1 and drug 2 requires that both drugs do not cause a DLT during the treatment cycle:
\begin{IEEEeqnarray*}{rCl}
1 - \pi_\perp(d_1, d_2) & = & \left(1 - \pi_1(d_1) \right) \left(1 - \pi_2(d_2) \right),
\end{IEEEeqnarray*}
where $d_1$ and $d_2$ refer to the doses of drugs 1 and 2, and $\pi_1$ and $\pi_2$ are the single-drug models for drugs 1 and 2, respectively.
The general case of $N$ drugs where $d_i$ refers to the dose of the $i$-th out of $N$ drugs follows directly from this as
\begin{IEEEeqnarray}{rCl}
1 - \pi_\perp(d_1, \dots, d_N) & = & \prod_{i=1}^{N} \left(1 - \pi_i(d_i) \right). \label{eq:independentModel}
\end{IEEEeqnarray}
This obviously does not take into account any drug-drug interactions yet.

\subsection{Drug-drug interaction models}
In practice, however, the DLT probability $\pi_{\bm{\eta}}(d_1,\ldots,d_N)$ for a combination of $N$ drugs may exceed the independence model,
\begin{IEEEeqnarray*}{rCl}
 \pi_{\bm{\eta}}(d_1,\ldots,d_N) &>& \pi_{\perp}(d_1,\ldots,d_N),
\end{IEEEeqnarray*}
if the drugs exhibit \emph{synergistic} toxic effects, or alternatively, under \emph{antagonistic} toxicity, we may have 
\begin{IEEEeqnarray*}{rCl}
 \pi_{\bm{\eta}}(d_1,\ldots,d_N) &<& \pi_{\perp}(d_1,\ldots,d_N).
\end{IEEEeqnarray*}
Note that in the general $N$-combination case, the set $\mathcal I_N$ of possible interactions consists of all pairwise and higher-order interactions:
\begin{IEEEeqnarray*}{rCl}
 \mathcal I_N &=& \mathcal P\bigl( \{ 1,\ldots,N\} \bigr) \setminus \bigcup_{i = 1}^N \{i\} \setminus \emptyset,
\end{IEEEeqnarray*}
where $\mathcal P$ is the power set. Accordingly, if we model each drug-drug interaction with a single parameter $\eta$, then the parameter vector $\bm{\eta} = (\eta_1, \ldots, \eta_{K_N})$ has dimension $K_N = \left| \mathcal I_N \right| = 2^N - N - 1$. In the following, we will also use the notation $\eta_s$ (where $s$ is a set of drugs denoted by their respective index $i$) to refer to the entry of $\bm{\eta}$ corresponding to the interaction between the drugs contained in set $s$.
\subsubsection{Desirable properties} \label{sec:desirableProperties}
Our goal is to incorporate both single-agent and combination data in a joint model $\pi_{\bm{\eta}}$. To this end, we will define desirable properties for the model and expand upon previously-published conditions \cite{Thall2003, Neuenschwander2014} for the drug-drug interaction model as well as generalize them to arbitrary interaction order. In particular, we require the following properties that ensure composability, i.e., compatibility of the multi-drug model with the single-drug models (or $N$-drug models with $(N-1)$-drug models in general), and with a model where DLTs are independently generated by each drug:
\begin{enumerate}
    \item The multi-drug toxicity model $\pi_{\bm{\eta}}(d_1, \dots, d_N)$ with drug-drug interaction parameters $\bm{\eta}$ reduces to the model without drug $i$ if the dose of drug $i$ is 0 -- i.e., the model reduces to the equivalent lower-order multi-drug model. For example, the two-drug model should fulfill, 
    \begin{IEEEeqnarray*}{rCl}
    	\pi_{\eta}(d_1, 0) & \mbeq & \pi_1(d_1) \text{  and  } \pi_{\eta}(0, d_2) \mbeq \pi_2(d_2).
    \end{IEEEeqnarray*}    
    More generally, for N drugs, we require
    \begin{IEEEeqnarray}{rCl}
    	\pi_{\bm{\eta}}(d_1, \dots, d_i = 0, \dots, d_N) & \mbeq & \pi_{\bm{\eta}}(d_1, \dots, d_{i-1}, d_{i+1}, \dots, d_{N}) \label{eq:noContributionAtZeroDose}
    \end{IEEEeqnarray} 
    for any $i \in [1, ..., N]$.
    \item The model reduces to $\pi_\perp$ in the case where all drugs generate DLT events independently, i.e., when the interaction strength parameter $\bm{\eta} = \bm{\eta}_\perp$. Without loss of generality, $\bm\eta_\perp = \bm 0$. For example, the two-drug model should fulfill, 	
	\begin{IEEEeqnarray*}{rCl}
    	\pi_{0}(d_1, d_2) & \mbeq & \pi_\perp(d_1, d_2).
    \end{IEEEeqnarray*}
    More generally, for N drugs, we require
    \begin{IEEEeqnarray}{rCl}
    	\pi_{\bm{0}}(d_1, \dots, d_N) & \mbeq & \pi_\perp(d_1, \dots, d_{N}).  \label{eq:reducesToIndependent}
    \end{IEEEeqnarray} 
\end{enumerate}
In addition to composability, we would like to have the following:
\begin{enumerate}
	\setcounter{enumi}{2}
    \item If any drug dose goes to infinity, the probability of a DLT should approach one, regardless of the other drug doses and  interaction strengths. This property formalizes the intuitive concept of "the dose makes the poison", credited to Paracelsus \cite{Grandjean2016}. I.e., at high doses every substance will become toxic -- this even holds for water \cite{ROWNTREE1923, Gardner2002}. For example, the two-drug model should fulfill
    	\begin{IEEEeqnarray*}{rCl}
    	\lim_{d_1 \to \infty} \pi_{\eta}(d_1, d_2) & \mbeq & 1 \forall d_2 \text{ and } \lim_{d_2 \to \infty} \pi_{\eta}(d_1, d_2) \mbeq 1 \forall d_1
    \end{IEEEeqnarray*}
    for any $\eta$. More generally, for N drugs, for any choice of $\bm{\eta}$ and any drug $i$, we require
    \begin{IEEEeqnarray}{rCl}
    \forall i \in \left\{1, ..., N\right\}: \lim_{d_i \to \infty} \pi_{\bm{\eta}}(d_1, \dots, d_i, \dots, d_N) & \mbeq & 1. \label{eq:asymptoticToxicity}
    \end{IEEEeqnarray} 
    \item We would like to be able to model both synergistic and antagonistic toxicities. That is, if drugs 1 and 2 exhibit synergistic toxicity when combined, the combined toxicity probability should be strictly higher than what one would expect if A and B were independently generating DLT events:
    \begin{IEEEeqnarray*}{rCll}
    	 \pi_{\eta}(d_1, d_2) & > & \pi_{\perp}(d_1, d_2) & \hspace{1em}\text{when }\eta > 0.
    \end{IEEEeqnarray*}
    Conversely, if drugs A and B exhibit antagonistic toxicity when combined, the combined toxicity probability should be strictly lower than what one would expect if A and B were independently generating DLT events:
    \begin{IEEEeqnarray*}{rCll}
    	 \pi_{\eta}(d_1, d_2) & < & \pi_{\perp}(d_1, d_2) & \hspace{1em}\text{when }\eta < 0.
    \end{IEEEeqnarray*}
    \item Last, but not least, for drugs that exhibit antagonistic toxicity, a desired property is to be able to model locally non-monotonic dose-responses. A classical example here would be treatment of methanol intoxication by administering ethanol: methanol and ethanol in combination are less toxic than methanol alone with respect to inducing blindness\cite{Marraffa2012} (at least up to a point; see property \eqref{eq:asymptoticToxicity}). More formally, for any finite choice of $d_1 > 0, d_2 > 0$ and $m_1>0, m_2 > 0$ with $0<\epsilon_1 <m_1, 0<\epsilon_2<m_2$, a choice of parameters exists such that toxicity is locally decreasing and we can find
    \begin{IEEEeqnarray*}{rCl}
	    \pi_{\eta}(d_1 + \epsilon_1, d_2 + \epsilon_2) & < & \pi_{\eta}(d_1, d_2). 
    \end{IEEEeqnarray*}

    More generally, if the drugs in set $s$ exhibit a synergistic or antagonistic toxicity, while all other drug combinations do not interact / act independently, we would like to enforce
    \begin{IEEEeqnarray}{rCl}
    	 \pi_{\bm{\eta}}(\bm{d}) & > & \pi_{\perp}(\bm{d}) \text{ where } \eta_s > 0, \text{ or }  \label{eq:synergisticToxicity} \\
    	 \pi_{\bm{\eta}}(\bm{d}) & < & \pi_{\perp}(\bm{d}) \text{ where } \eta_s < 0, \text{ respectively, with} \label{eq:antagonisticToxicity} \\
    	 \bm{\eta} & = &  \left(0, \dots, 0, \eta_s, 0, \dots, 0\right)^\intercal, s \in \mathcal I_N. \IEEEnonumber
    \end{IEEEeqnarray}
    In addition, for antagonistically-toxic cases, we require the model be locally non-monotonic (again, if all other drug combinations interact independently). That is, for any finite choice of $ d_{\bm{s}} > \bm{0}, \bm{m}>\bm{0}, 0<\epsilon_i<m_i, i \in s$,  a choice of parameters exists such that toxicity is locally decreasing and we can find
    \begin{IEEEeqnarray}{rCl}
	    \pi_{\bm{\eta}}(\bm{d_{s}} + \bm{\epsilon_s}, \bm{d_{s^\complement}}) & < & \pi_{\bm{\eta}}(\bm{d_{s}}, \bm{d_{s^\complement}}) \label{eq:antagonisticToxicityNonMonotonic}
    \end{IEEEeqnarray}
\end{enumerate}

\subsubsection{Prior work}
The two-drug model proposed by Thall \textit{et al}\cite{Thall2003} 
\begin{IEEEeqnarray}{rCl}
\text{logit}\left(\pi_{\text{Thall}, \eta}(d_1, d_2)\right) & = & \log\left(\alpha_1 \left(\frac{d_1}{d_1^*}\right)^{\beta_1} + \alpha_2 \left(\frac{d_2}{d_2^*}\right)^{\beta_2} + \alpha_3\left( \left(\frac{d_1}{d_1^*}\right)^{\beta_1} \left(\frac{d_2}{d_2^*}\right)^{\beta_2} \right)^{\beta_3}\right) \\  & \text{ with } & \alpha_i \geq 0, \beta_i \geq 0 \\
& \text{ and } & \bm{\eta} = \left(\alpha_3, \beta_3\right) \label{eq:thall2003} 
\end{IEEEeqnarray} 
fulfills \eqref{eq:noContributionAtZeroDose} through \eqref{eq:antagonisticToxicity}. However, it does not support modelling non-monotonic dose-responses due to antagonistic toxicity interactions and therefore violates property \eqref{eq:antagonisticToxicityNonMonotonic} (though extensions based on the Generalized Aranda-Ordaz (GAO) model can remedy this, depending on the choice of function in the exponent). As a simple solution, we propose a model according  the previously suggested form\cite{Gasparini2010}
\begin{IEEEeqnarray}{rCl}
\text{logit}\left(\pi_{\bm{\eta}}(\bm{d})\right) & = & \text{logit}\left(\pi_\perp(\bm{d})\right) + \text{logit}\left(\pi_{\text{inter}, \bm{\eta}}(\bm{d})\right), \label{eq:logitsum}
\end{IEEEeqnarray} 
though, in contrast to Gasparini \emph{et al.}\cite{Gasparini2010}, here we do not allow the joint probability of toxicity to approach 1 or 0 with no restrictions, but rather impose the conditions mentioned in section~\ref{sec:desirableProperties}: \eqref{eq:logitsum} fulfils condition \eqref{eq:reducesToIndependent} if 
\begin{equation*}
\text{logit}\left(\pi_{\text{inter}, \bm{\eta}}(\bm{d})\right) = 0 \text{ for } \bm{\eta} = \bm{0}.
\end{equation*}

For condition \eqref{eq:noContributionAtZeroDose}, we then have
\begin{IEEEeqnarray*}{rCl}
\text{logit}\left(\pi_{\bm{\eta}}(d_1, \dots, d_i = 0, \dots, d_N)\right) & = & \text{logit}\left(\pi_\perp(d_1, \dots, d_i = 0, \dots, d_N)\right) + \text{logit}\left(\pi_{\text{inter}, \bm{\eta}}(d_1, \dots, d_i = 0, \dots, d_N)\right) \\
& = & \text{logit}\left(1 - \left(\left(1 - \pi_i(d_i = 0)\right) \prod_{j = 1, j \neq i}^{N} \left(1 - \pi_j(d_j) \right)\right)\right) + \text{logit}\left(\pi_{\text{inter}, \bm{\eta}}(d_1, \dots, d_i = 0, \dots, d_N)\right) \\
& = & \text{logit}\left(\pi_\perp(d_1, \dots, d_{i-1}, d_{i+1}, \dots, d_N)\right) + \text{logit}\left(\pi_{\text{inter}, \bm{\eta}}(d_1, \dots, d_i = 0, \dots, d_N)\right),
\end{IEEEeqnarray*} 
i.e., if the dose of drug $i$ is 0, the independent probability is the probability without the drug, and condition \eqref{eq:noContributionAtZeroDose} holds iff $\text{logit}\left(\pi_{\text{inter}, \bm{\eta}}(d_1, \dots, d_i = 0, \dots, d_N)\right) = \text{logit}\left(\pi_{\text{inter}, \bm{\eta}}(d_1, \dots, d_{i-1} , d_{i+1}, \dots, d_N)\right)$.
An established definition\cite{Neuenschwander2014} for $\text{logit}\left(\pi_{\text{inter}, \bm{\eta}}\right)$ that fulfils this condition is:
\begin{IEEEeqnarray}{rCl}
 \text{logit}\left(\pi_{\text{inter, linear}, \bm{\eta}}(d_1, \dots, d_N)\right) & = & \sum_{s \in \mathcal{I_N}} \eta_s \gamma_\text{linear}(\bm{d_s}), \text{ with }  \IEEEnonumber\\
 \gamma_\text{linear}(\bm{d_s}) & = &  \prod_{i \in s} \frac{d_i}{d_{i}^*}. \label{eq:gamma_linear}
\end{IEEEeqnarray}
However, it can be shown that this model violates condition \eqref{eq:asymptoticToxicity} when antagonistic toxicities ($\eta_s < 0$) are modelled. This is due to the interaction growing exponentially faster with dose than the contributions of the single drug components. For instance, in the two-drug-case, this causes $\lim_{d_1, d_2 \rightarrow \infty} \pi_{\eta < 0}(d_1, d_2) \rightarrow 0$, i.e., toxicity asymptotically decreases to zero as doses increase to infinity: an obviously undesirable property.
\subsubsection{An improved drug-drug interaction model}
To ensure that Paracelsus' principle holds (any drug approaching infinity leads to an event rate of 1), we suggest a novel interaction model that saturates the interaction strength with dose, which ensures that the single-drug toxicity will always dominate the overall toxicity contribution eventually:
\begin{IEEEeqnarray}{rCl}
 \text{logit}\left(\pi_{\text{inter, saturating}, \bm{\eta}}(d_1, \dots, d_N)\right) & = & \sum_{s \in \mathcal{I_N}} \eta_s \gamma_\text{saturating}(\bm{d_s}), \text{ with} \IEEEnonumber \\
 \gamma_\text{saturating}(\bm{d_s}) & = & \frac{2 \prod_{i \in s} \frac{d_i}{d_{i}^*}}{1 + \prod_{i \in s} \frac{d_i}{d_{i}^*}} = 2 \, \text{logit}^{-1}\left(\sum_{i \in s} \log\frac{d_i}{d_{i}^*} \right).  \label{eq:gamma_saturating}
\end{IEEEeqnarray}

\begin{center}
\begin{table}[th]%
\caption{Summary of interaction models and their properties. \label{tab:modelProperties}}
\centering
\begin{tabular*}{450pt}{@{\extracolsep\fill}rcccc@{\extracolsep\fill}}
\toprule
& \multicolumn{4}{@{}c@{}}{\textbf{Model}}  \\
\cmidrule{2-5}
\textbf{Property} & \textbf{No interaction} \eqref{eq:independentModel}  & \textbf{Thall \emph{et al}, 2003}\cite{Thall2003}   & \textbf{Linear} \eqref{eq:gamma_linear} & \textbf{Saturating} \eqref{eq:gamma_saturating}  \\
\midrule
\makecell[r]{Model reduces to lower-order \\ model at 0 dose \eqref{eq:noContributionAtZeroDose}} & \checkmark            &  \checkmark  &  \checkmark    &    \checkmark \vspace{0.2cm}  \\
\makecell[r]{Model compatible with \\ no-interaction model \eqref{eq:reducesToIndependent}} & \checkmark  & \checkmark  & \checkmark & \checkmark  \vspace{0.2cm}\\
\makecell[r]{$\pi \rightarrow 1$ as $d \rightarrow \infty$\eqref{eq:asymptoticToxicity}} &  \checkmark   &  \checkmark         & \xmark      &  \checkmark  \vspace{0.2cm}          \\
\makecell[r]{Can model synergistic \\ drug-drug interactions \eqref{eq:synergisticToxicity}} &   \xmark   & \checkmark & \checkmark & \checkmark  \vspace{0.2cm} \\
\makecell[r]{Can model antagonistic \\ drug-drug interactions \eqref{eq:antagonisticToxicity}} &   \xmark   & \checkmark & \checkmark & \checkmark  \vspace{0.2cm} \\
\makecell[r]{Can model non-monotonic \\ antagonistic drug-drug \\ interactions \eqref{eq:antagonisticToxicityNonMonotonic}} &   \xmark   & \xmark & \checkmark & \checkmark  \\
\bottomrule
\end{tabular*}
\end{table}
\end{center}

This model is the only one of the above that satisfies conditions \eqref{eq:noContributionAtZeroDose}-\eqref{eq:antagonisticToxicityNonMonotonic} (see table~\ref{tab:modelProperties}). Note that, for both interaction models \eqref{eq:gamma_linear} and \eqref{eq:gamma_saturating}, it holds that $\gamma_\text{linear}\left(\bm{d}^*\right) = \gamma_\text{saturating}\left(\bm{d}^*\right) = 1$, but in contrast to $\gamma_\text{linear}$, $\gamma_\text{saturating}$ will saturate at 2 if any dose goes to infinity.

\section{Results}
We will now assess how above models perform under different scenarios, i.e., different hypothetical ground truths, by illustrating their behavior when sequentially observing pre-defined  cohorts of hypothetical data. These data scenarios allow for a deep look at the model behavior, which is critical to understand the models' performance under hypothetical truths that are of particular interest by a clinical team, and we use data scenarios to give an intuition for the interaction models here.

\subsection{Model and prior setup}
We will investigate each model with a prior set to 10\% mean toxicity at the reference dose of 200 (which will be a dose we observe combination data at, for simplicity), the interaction parameter $\eta$ for the linear interaction model  $\pi_{\text{inter, linear}, \bm{\eta}}$ and the saturating interaction model $\pi_{\text{inter, saturating}, \bm{\eta}}$ centered at 0 (no interaction), and the other parameter priors chosen (corresponding to previous work\cite{Neuenschwander2014}) as
\begin{equation}
\log \alpha \sim \mathcal{N}\left(\text{logit}(0.10),\,2^{2}\right), \quad
\log \beta \sim \mathcal{N}\left(0,\,1^2\right), \quad
\eta \sim \mathcal{N}\left(0,\,\sigma_\text{inter}^2\right).
\end{equation} 
For $\pi_{\text{Thall}, \bm{\eta}}$, we choose
\begin{equation}
\log \alpha_3 \sim \mathcal{N}\left(2~\text{logit}(0.10),\,\sigma_{\alpha_3}^2\right), \quad
\log \beta_3 \sim \mathcal{N}\left(0,\,\sigma_{\beta_3}^2\right).
\end{equation} 
We will investigate four model variants that vary the interaction parameters:
\begin{enumerate}
\item The no-interaction model, $\pi_{\perp}$,
\item The model proposed in Thall \emph{et al}\cite{Thall2003}, $\pi_{\text{Thall}, \bm{\eta}}$, with $\left(\sigma_{\alpha_3}, \sigma_{\beta_3}\right) = \left(0.5 \sqrt{2~2^2}, 0.5\right) $ and $\left(\sqrt{2~2^2}, 1\right)$ for the interaction term,
\item the linear interaction model\cite{Neuenschwander2014}  $\pi_{\text{inter, linear}, \bm{\eta}}$ with $\sigma_\text{inter} = 0.5$ (a narrow prior on the interaction strength) and $\sigma_\text{inter} = 1.5$ (a wider prior), as well as
\item the novel saturating interaction model $\pi_{\text{inter, saturating}, \bm{\eta}}$ with $\sigma_\text{inter} = 0.5\text{ and }1.5$.
\end{enumerate}
To focus the data scenarios onto the differences in the model structure, we assume that any historical data can be pooled with the ongoing trial data, and forego modelling between-trial heterogeneity. In practice, one would only want to do this if the single drug data comes from the same trial and population, and account for between-trial heterogeneity otherwise as previously described\cite{Neuenschwander2016}.

\subsection{Data scenarios}
We investigate data scenarios (summarized in table~\ref{tab:dataScenarios}) for a combination trial of drugs A and B with historical data for the individual drugs A and B, and take a deeper look at potential outcomes and the resulting response surface after different hypothetical first 5-patient combination cohorts. Here, we achieve the following (practical) goals for the model and model parametrization:

\begin{center}
\begin{table}[th]%
\caption{Summary of data scenarios. \label{tab:dataScenarios}}
\centering
\begin{tabular*}{400pt}{@{\extracolsep\fill}rcrrrr@{\extracolsep\fill}}
\toprule
& \multicolumn{5}{@{}c@{}}{\textbf{First combination cohort}}  \\
\cmidrule{1-6}
\textbf{Figure} & \textbf{Historical data}${}^\dagger$  & \textbf{Drug 1 dose (mg)}   & \textbf{Drug 2 dose (mg)}  & \textbf{\# Patients} & \textbf{\# DLT events}   \\
\midrule
\ref{fig:prior_histdata}AB & \xmark            &    &      &   &    \\
\ref{fig:prior_histdata}CD  &  \checkmark      &    &      &   &    \\
\ref{fig:prior_histdata_0_5}  &  \checkmark    & 200 & 200 & 5 & 0  \\
\ref{fig:prior_histdata_5_5}AB  &  \checkmark  & 200 & 200 & 5 & 5  \\
\ref{fig:prior_histdata_5_5}CD  &  \checkmark  & 100 & 100 & 5 & 5  \\
\bottomrule
\end{tabular*}
\begin{tablenotes}
\item[${}^\dagger$] As defined in table~\ref{tab:histDataSingleDrug}.
\end{tablenotes}
\end{table}
\end{center}
\begin{center}
\begin{table}%
\caption{Single-drug DLT data summary for drugs A and B used as historical data for the data scenarios on the trial combining drugs A and B.\label{tab:histDataSingleDrug}}
\centering
\begin{tabular*}{400pt}{@{\extracolsep\fill}rrrrrrrr@{\extracolsep\fill}}
\toprule
& & \multicolumn{6}{@{}c@{}}{\textbf{\# DLT events at dose (mg)}}  \\
\cmidrule{3-8}
\textbf{Drug} & \textbf{\# Patients per dose} & \textbf{50}  & \textbf{100}  & \textbf{200}  & \textbf{300} & \textbf{400} & \textbf{600}  \\
\midrule
1 & 10  & 0  & 1  & 1  & 2 & 3 & 6  \\
2 & 5   & 0  & 0  & 1  & 1 & 1 & 3 \\
\bottomrule
\end{tabular*}
\end{table}
\end{center}

\begin{figure}[p]
\centering
{
\begin{subfigure}[t]{.45\linewidth}
\caption{}
{\includegraphics[width=240pt]{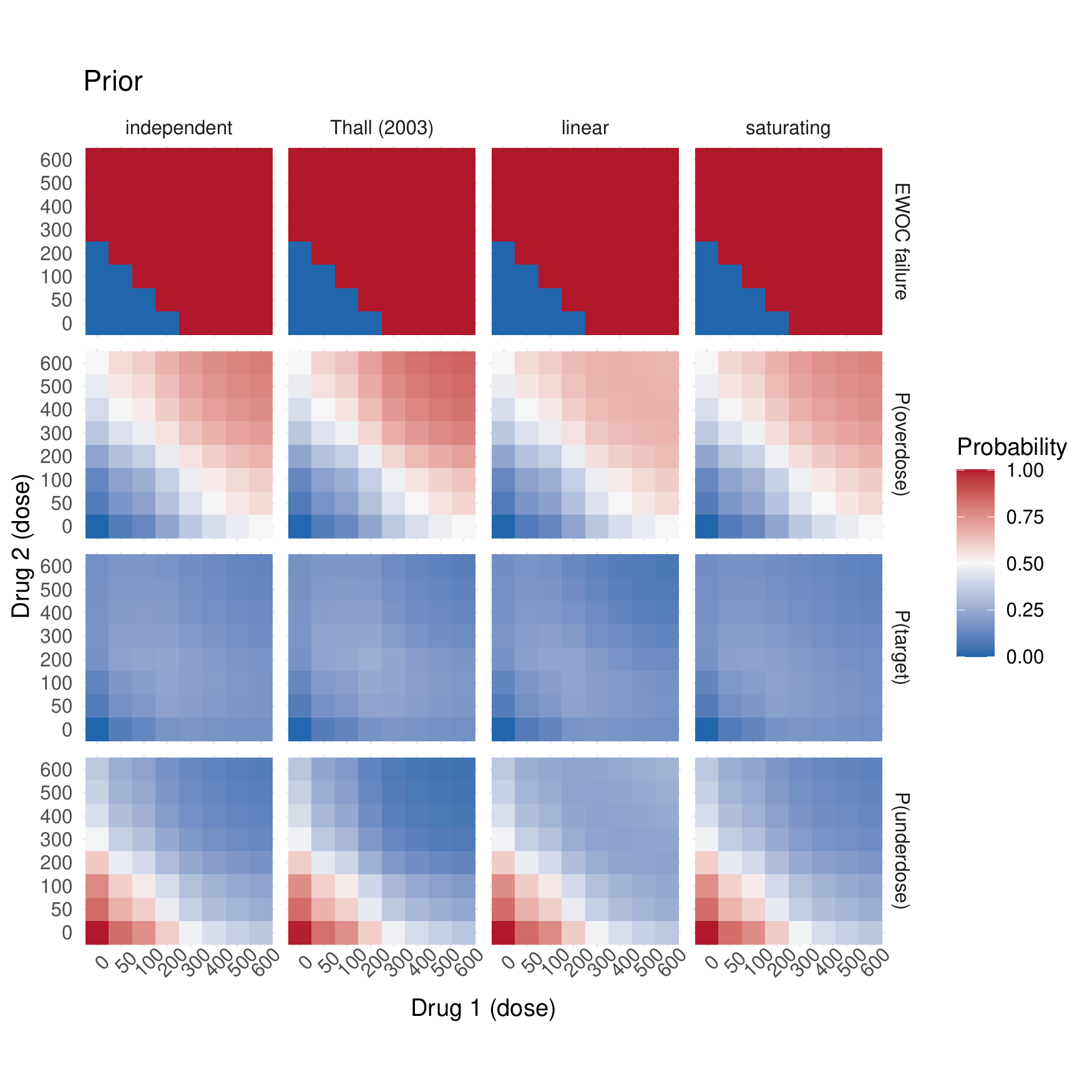}}
\end{subfigure}
\hspace{1mm}
\begin{subfigure}[t]{.45\linewidth}
\caption{}
{\includegraphics[width=240pt]{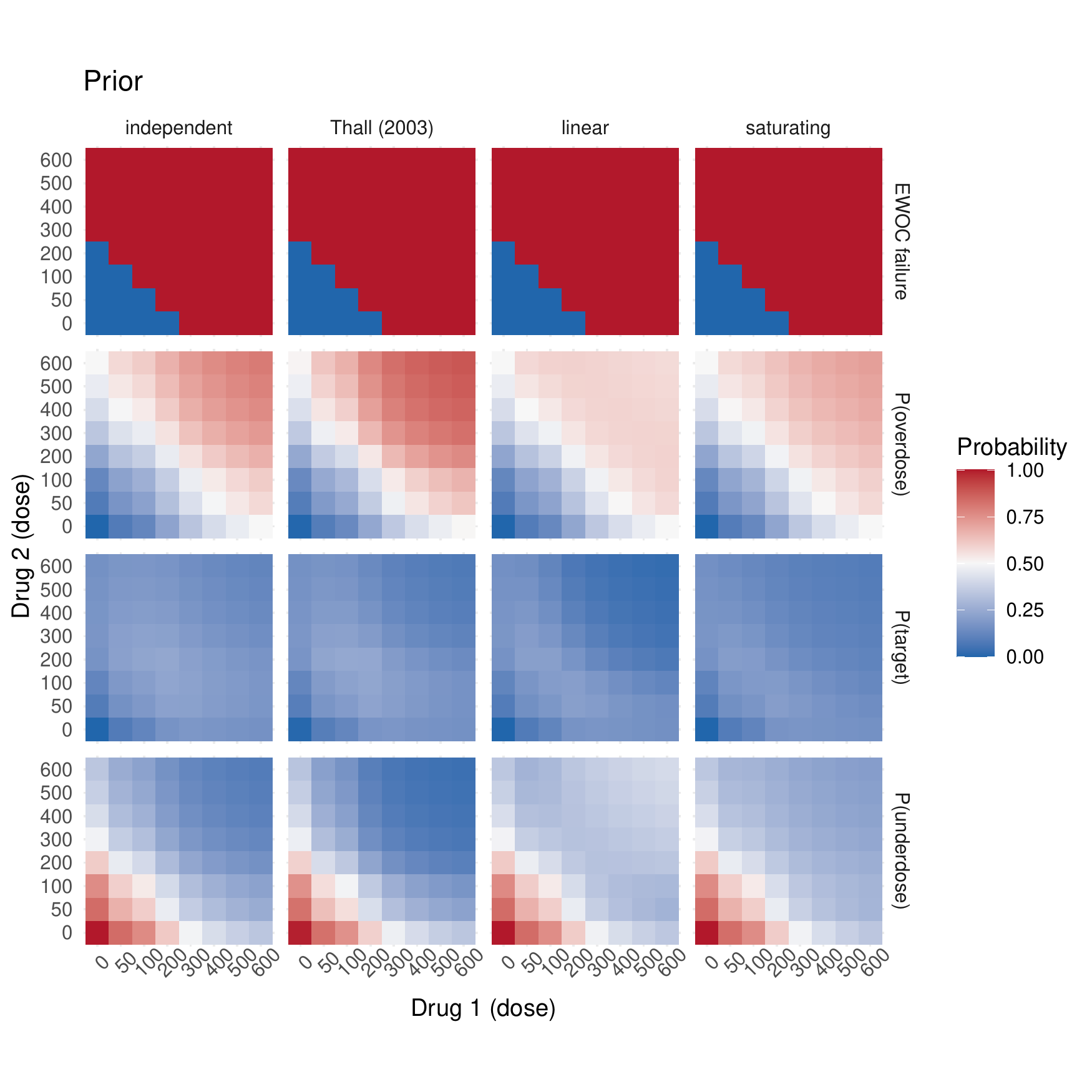}}
\end{subfigure}
\\
\begin{subfigure}[t]{.45\linewidth}
\caption{}
{\includegraphics[width=240pt]{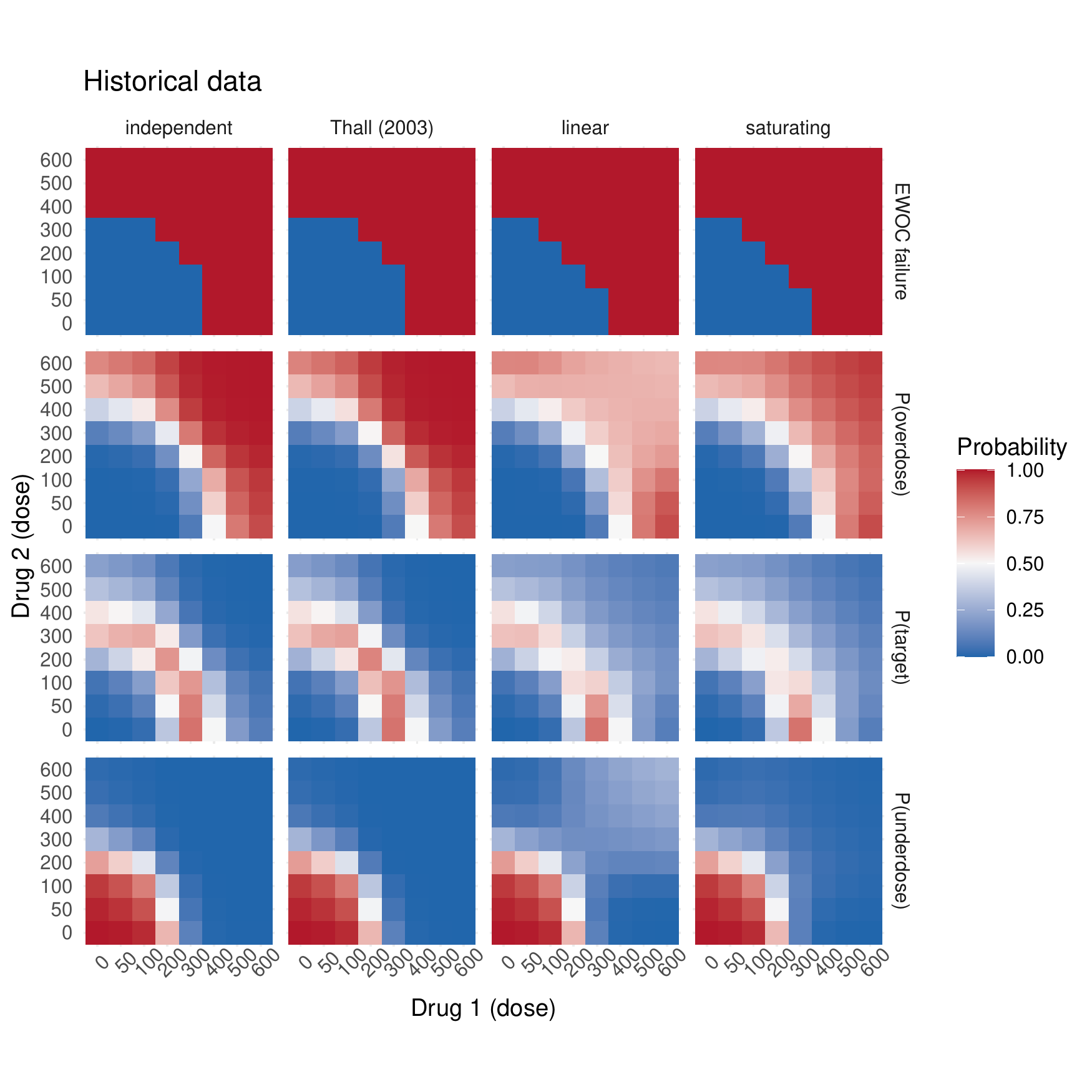}}
\end{subfigure}
\hspace{1mm}
\begin{subfigure}[t]{.45\linewidth}
\caption{}
{\includegraphics[width=240pt]{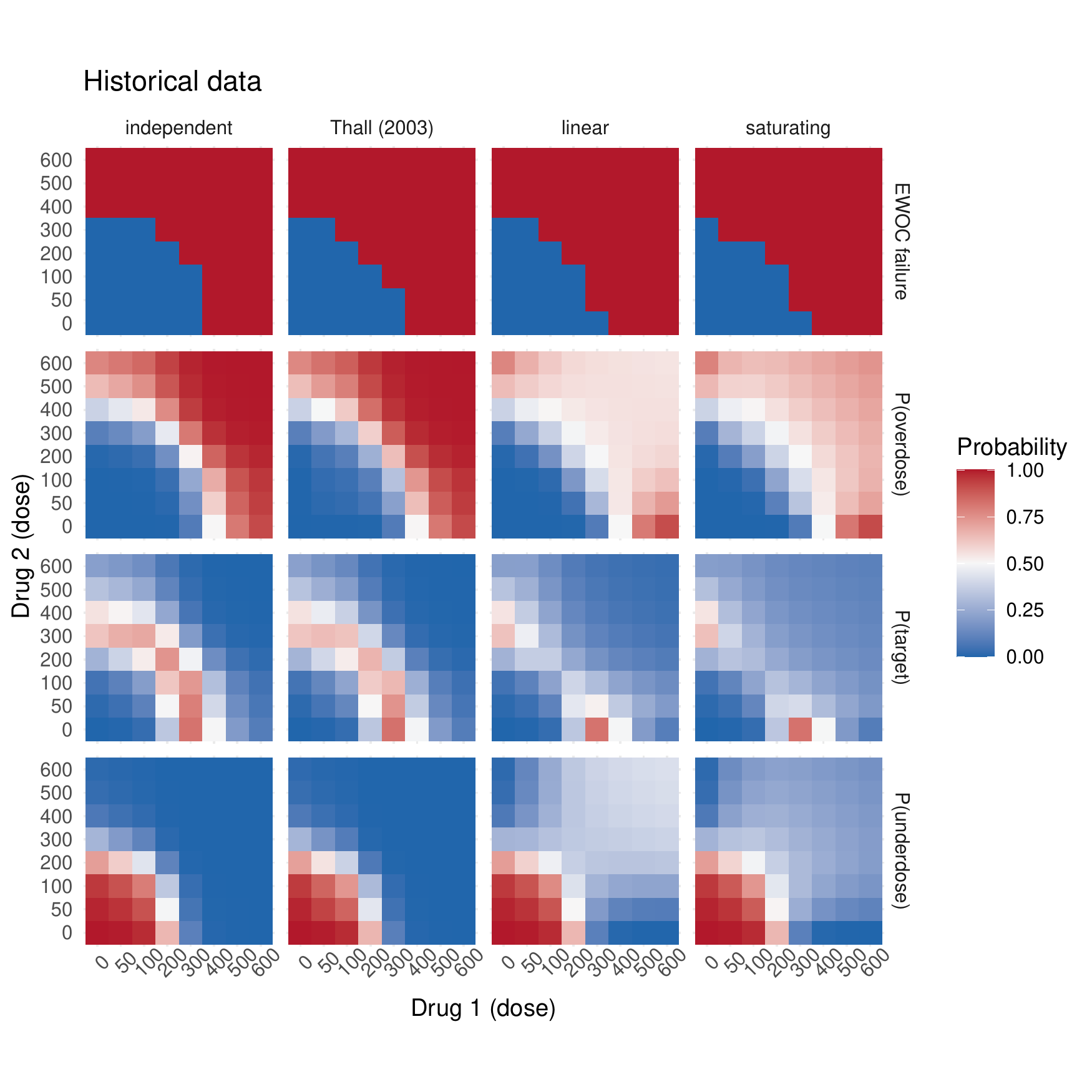}}
\end{subfigure}
}
\caption{Prior dose-response surface with \textbf{A} $\sigma_\text{inter} = 0.5,\sigma_{\alpha_3} = 0.5 \sqrt{2~2^2}, \sigma_{\beta_3} = 0.5$
, or \textbf{B} $\sigma_\text{inter} = 1.5, \sigma_{\alpha_3} = \sqrt{2~2^2}, \sigma_{\beta_3} = 1$. \textbf{C} and \textbf{D} show the dose-response surface after updating the models in \textbf{A} and \textbf{B} with the single-drug historical data for drugs 1 and 2, respectively.}
\label{fig:prior_histdata}
\end{figure}

\begin{figure}
\centering
{
\begin{subfigure}[t]{.45\linewidth}
\caption{}
{\includegraphics[width=240pt]{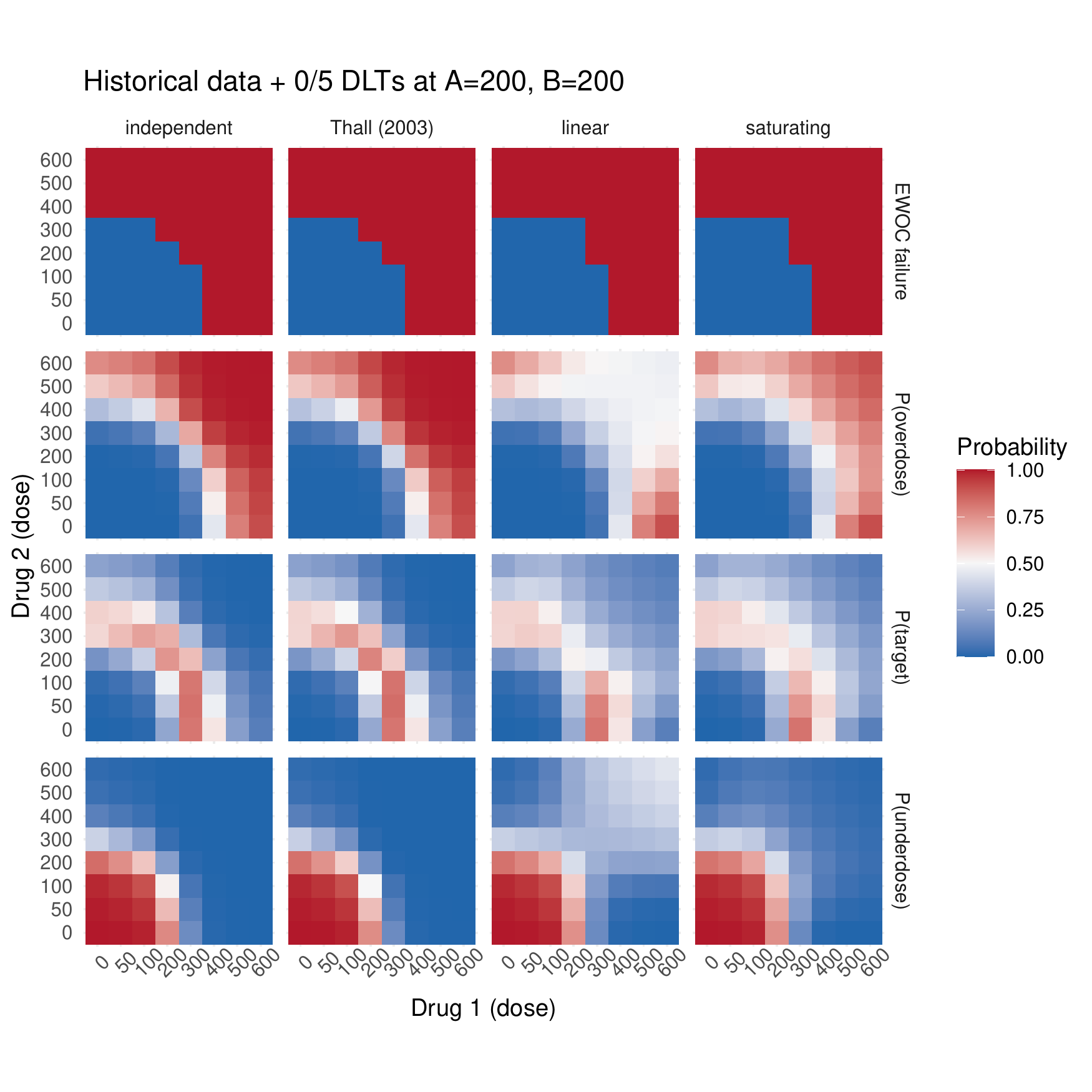}}
\end{subfigure}
\hspace{1mm}
\begin{subfigure}[t]{.45\linewidth}
\caption{}
{\includegraphics[width=240pt]{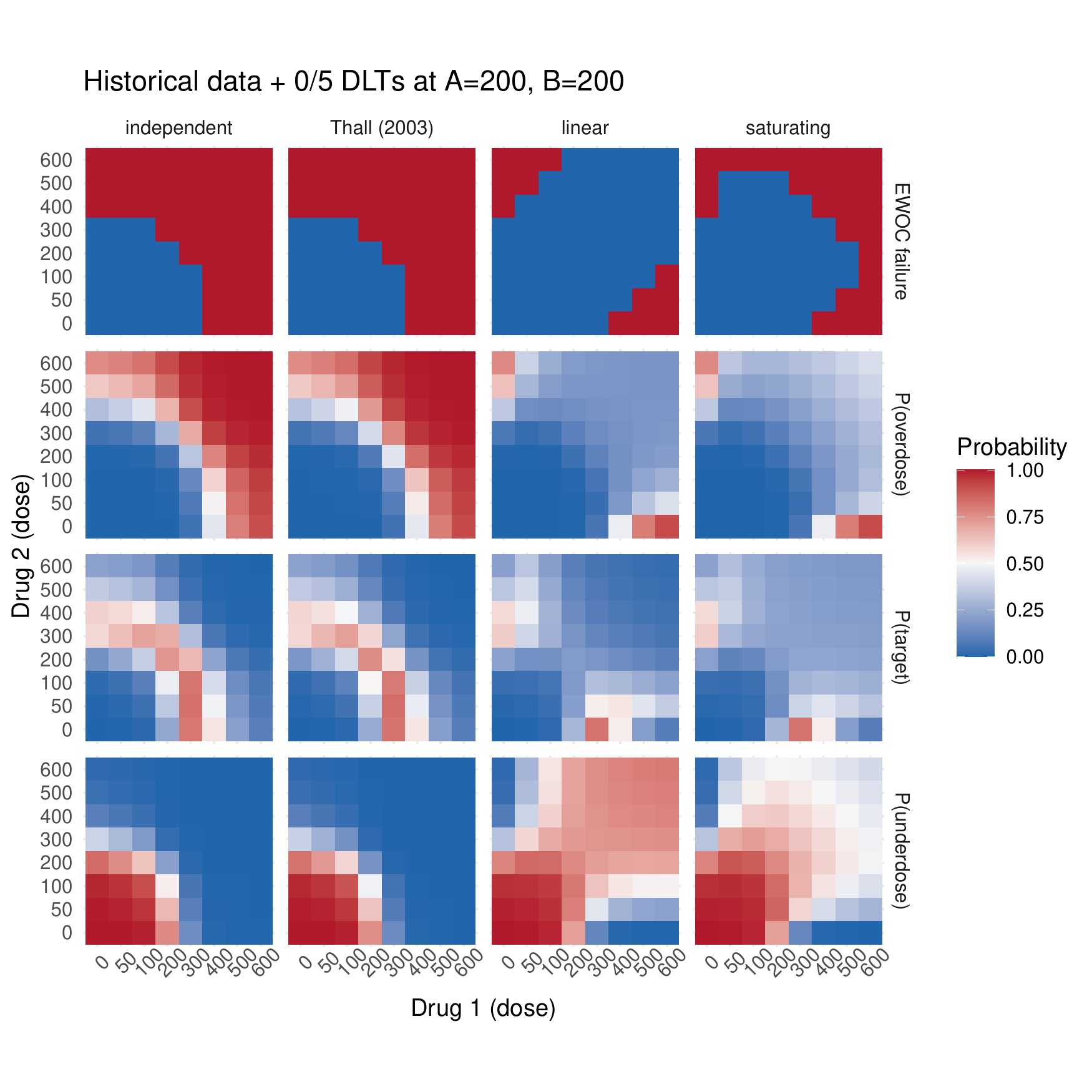}}
\end{subfigure}
}
\caption{Dose-response surface with \textbf{A} $\sigma_\text{inter} = 0.5,\sigma_{\alpha_3} = 0.5 \sqrt{2~2^2}, \sigma_{\beta_3} = 0.5$
, or \textbf{B} $\sigma_\text{inter} = 1.5, \sigma_{\alpha_3} = \sqrt{2~2^2}, \sigma_{\beta_3} = 1$ after updating the prior with historical data and a single combination cohort of 5 patients at a dose of 200 for drugs 1 and 2, respectively, with 0 DLT events.}
\label{fig:prior_histdata_0_5}
\end{figure}

\begin{figure}
\centering
{
\begin{subfigure}[t]{.45\linewidth}
\caption{} \label{fig:prior_histdata_5_5_200_05}
{\includegraphics[width=240pt]{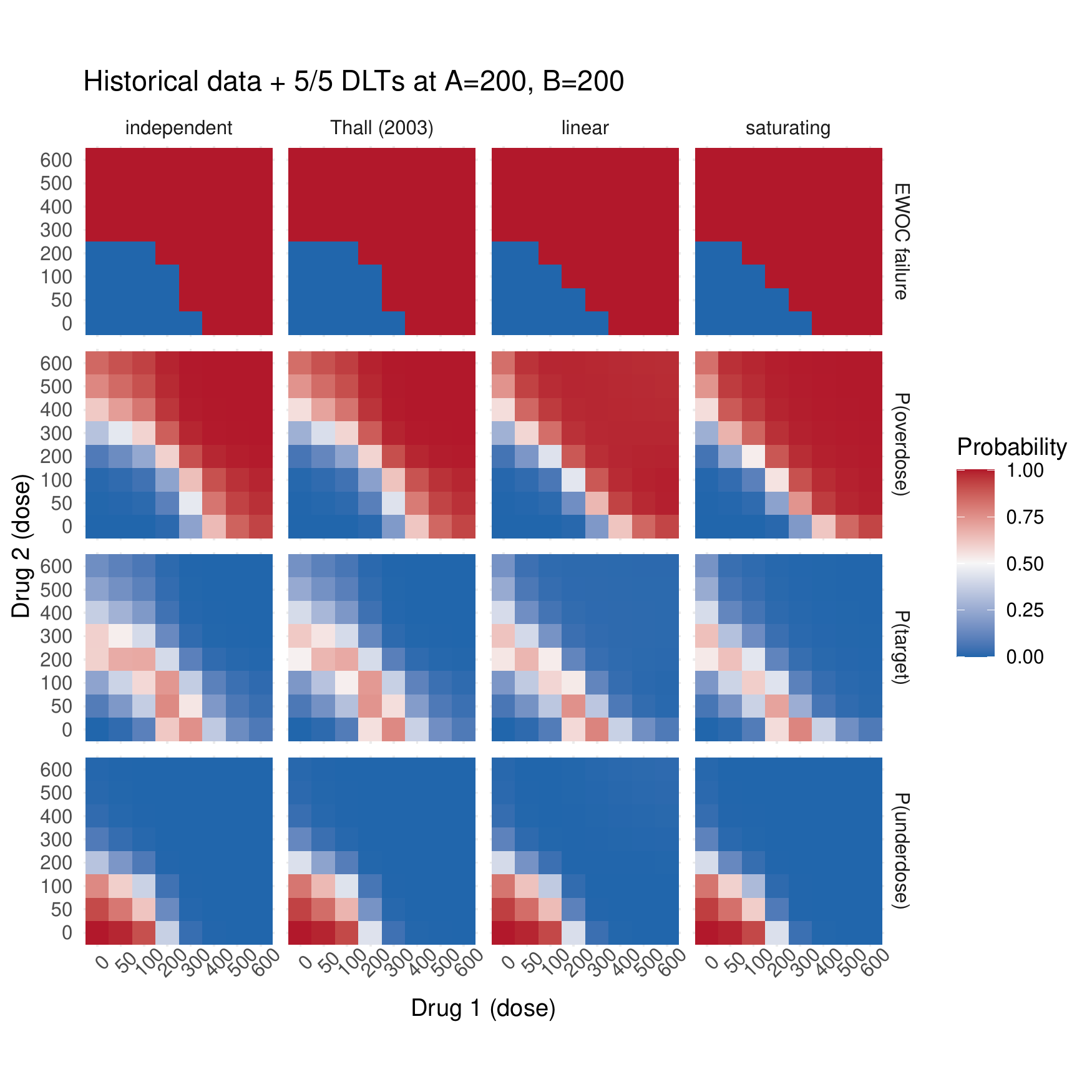}}
\end{subfigure}
\hspace{1mm}
\begin{subfigure}[t]{.45\linewidth}
\caption{} \label{fig:prior_histdata_5_5_200_15}
{\includegraphics[width=240pt]{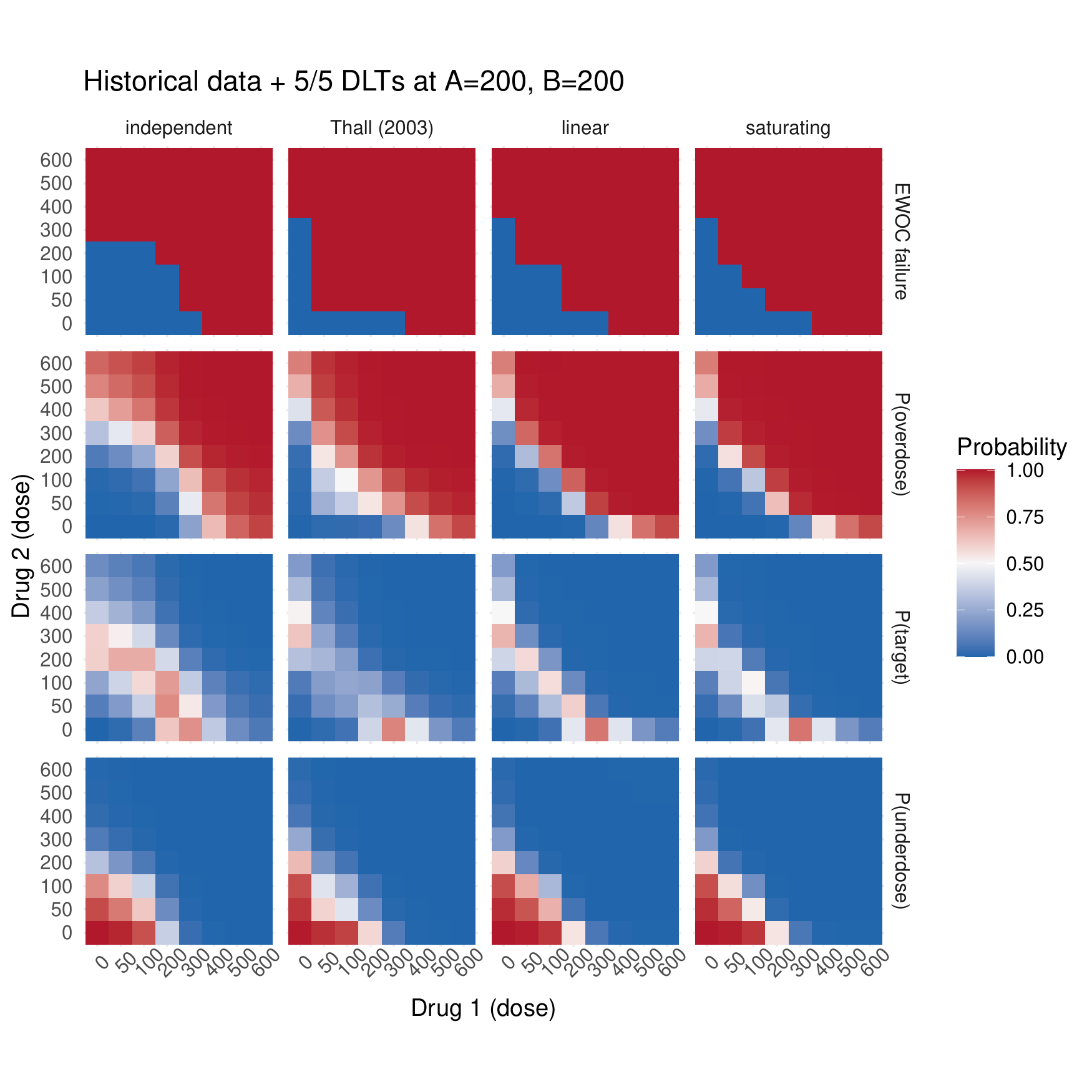}}
\end{subfigure}
\\
\begin{subfigure}[t]{.45\linewidth}
\caption{} \label{fig:prior_histdata_5_5_100_05}
{\includegraphics[width=240pt]{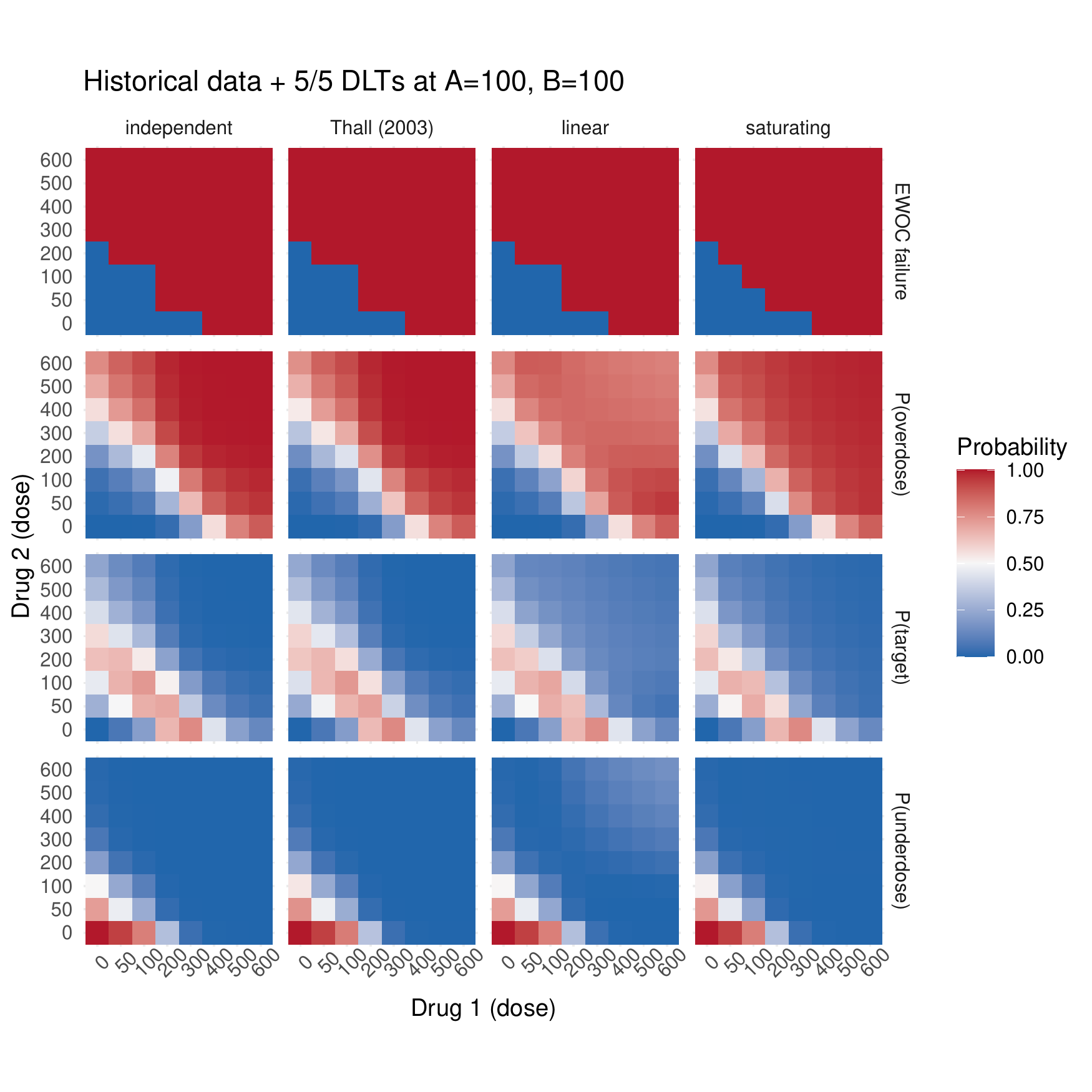}}
\end{subfigure}
\hspace{1mm}
\begin{subfigure}[t]{.45\linewidth}
\caption{} \label{fig:prior_histdata_5_5_100_15}
{\includegraphics[width=240pt]{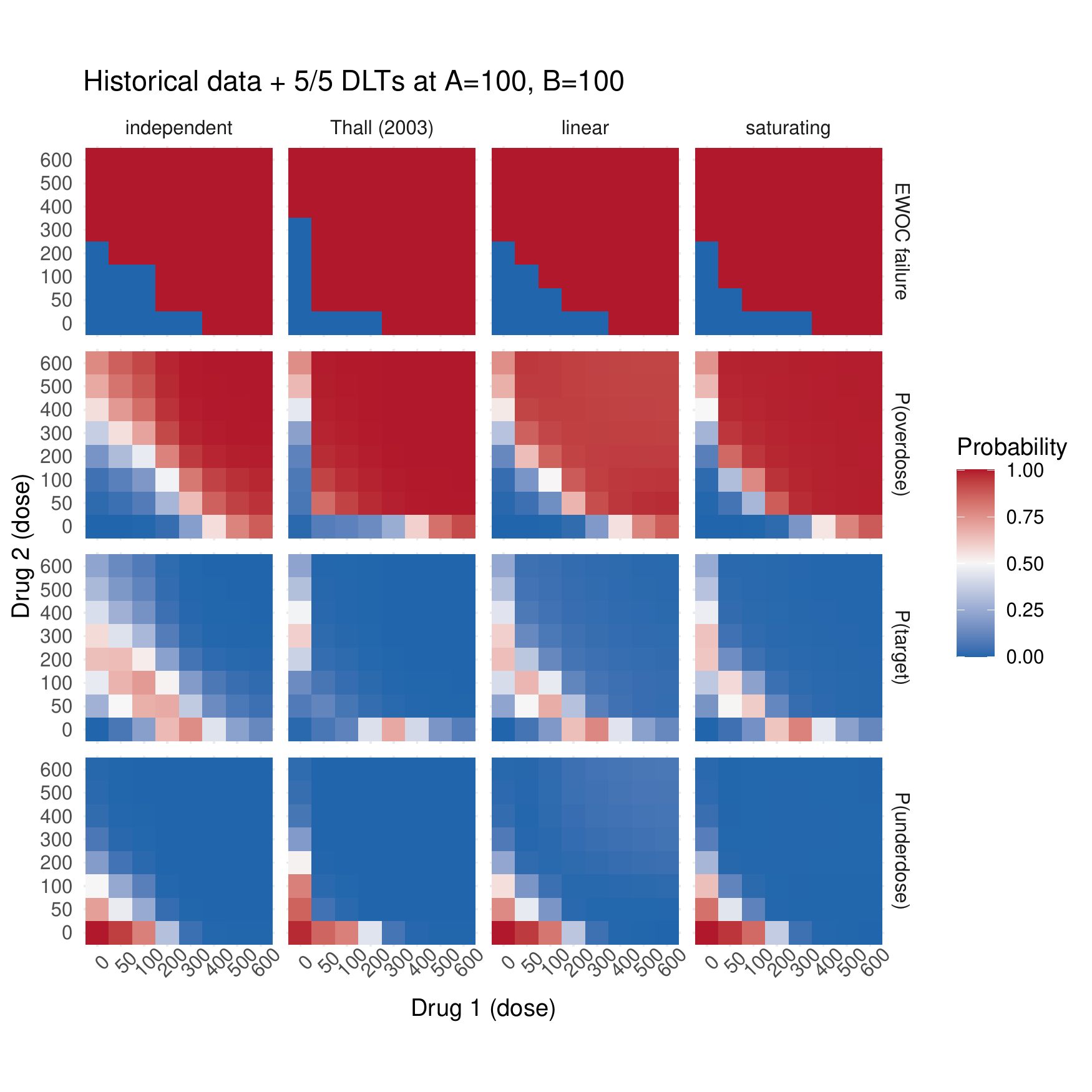}}
\end{subfigure}
}
\caption{Dose-response surface with (\textbf{A, C}) $\sigma_\text{inter} = 0.5,\sigma_{\alpha_3} = 0.5 \sqrt{2~2^2}, \sigma_{\beta_3} = 0.5$
, or (\textbf{B, D}) $\sigma_\text{inter} = 1.5, \sigma_{\alpha_3} = \sqrt{2~2^2}, \sigma_{\beta_3} = 1$ after updating the prior with historical data and a single combination cohort of 5 patients with 5 DLT events at a dose of 200 (\textbf{A, B}) or 100 (\textbf{C, D}) for drugs 1 and 2, respectively.}
\label{fig:prior_histdata_5_5}
\end{figure}

\begin{enumerate}
\item Before observing the first combination cohort, it should be possible to encode our expectations on the drug-drug interaction from, e.g., pre-clinical data into a reasonable prior. For instance, we could center our prior on no interaction while allowing for some variability towards synergistic and antagonistic interactions. On top of this, we obviously want to be able to take into account single-drug historical data (detailed in table~\ref{tab:histDataSingleDrug}). This is trivially achievable with our chosen model formulation, and the initial prior without historical data (Figures~\ref{fig:prior_histdata}~A and \ref{fig:prior_histdata}~B) can easily be updated with the single-drug historical data, in this case allowing for a larger range of drug A/B dose combinations that satisfy the EWOC criterion (Figures~\ref{fig:prior_histdata}~C and \ref{fig:prior_histdata}~D).

\item Leverage the information from the single 5-patient cohort for further decision making in the combination trial. In particular, there are two scenarios for the observed dose combination (with equal dose in A and B) where we would like to have the following properties:
\begin{enumerate}
\item If the outcome is safer than expected from the single-drug historical data if the drugs were generating DLT events independently, e.g., 0/5 DLTs at a dose combination of $A = B = 200$ -- indicating antagonistic toxicity -- we would like to further explore the region where both drugs are applied in a similar ratio. If the MTD is in this region, we would like to find it (or a dose close to it). However, we also would like to still exercise caution and not escalate too quickly in this region, and not have a model that allows us to arbitrarily escalate both drugs. This is precisely what the saturating drug-drug interaction model formulation presented here allows for (Figure~\ref{fig:prior_histdata_0_5}) -- while the linear formulation cannot: either we assume the interaction to be close to independent, preventing escalation to infinity but also precluding exploration of higher A/B combinations, or we assume a wider prior, and allow for escalation to infinity while also allowing for exploration higher A/B combinations -- achieving both (if possible at all) requires walking a very fine line.
\item If the outcome suggests high toxicity, e.g., 5/5 DLTs, we require the model to no longer allow this combination -- even if the toxic combination dose is relatively low compared to the acceptable single-drug dosing. To achieve this, the combination model must be sufficiently flexible, and the prior on having an independent interaction should not be too strong. We illustrate two such cases: 5/5 DLTs at 200/200 (Figure~\ref{fig:prior_histdata_5_5}AB), and a more extreme scenario with 5/5 DLTs at 100/100 (Figure~\ref{fig:prior_histdata_5_5}CD). While continuing on a dose of 200/200 is prevented by all models -- including the one without any interaction term and the models with low interaction parameter variability (Figure~\ref{fig:prior_histdata_5_5_200_05}), this is no longer true for the more extreme scenario (Figure~\ref{fig:prior_histdata_5_5_100_05}). After observing 5/5 DLTs, continuing at the same dose combination is undesirable -- modeling the synergistic drug-drug interaction between A and B with our proposed interaction term would prevent this.

\end{enumerate}
\item Largely preserve the marginals for drug A and B if the sample size for the individual drugs is substantially higher (such as in the example given in table~\ref{tab:histDataSingleDrug}, 60 patients for drug A, and 30 patients for drug B). That is, if the dose for drug A is $0$ (or conversely, the one for drug B is $0$), the prediction should closely resemble the one that does not consider the single 5-patient cohort of combination data. While this is the case for all models in the 0/5 DLT scenario (Figure~\ref{fig:prior_histdata_0_5}), this is no longer holds for the 5/5 DLT scenarios: at a respective dose of 200 for drugs A and B, not using any interaction model or the models with low interaction parameter variability leads to the marginal of drug B (which has half the historical data of drug A) no longer allowing for dose 300 (Figure~\ref{fig:prior_histdata_5_5_200_05}), while the interaction models with $\sigma_\text{inter} = 1.5$ manage to preserve the marginals for drug B (Figure~\ref{fig:prior_histdata_5_5_200_15}). To also prevent the marginals updating in a manner in which, for 5/5 DLTs at a drug A and B dose of 100 (respectively), a single-drug B dose of 300 is no longer allowed, one could choose slightly stronger priors on the single-drug dose-response curve.
\end{enumerate}

\section{Discussion}\label{sec:discussion}
Drug-drug interactions in Oncology have received substantial attention over the last two decades \cite{Beijnen2004}, and ways of systematically addressing them when modelling dose-dependent DLT probability in phase-I Oncology have been established \cite{Thall2003}. In contrast to the recently-proposed notion that ignoring drug-drug interactions when modelling safety in phase-I Oncology combination trials is beneficial in general \cite{Mozgunov2021}, we think that this depends on the type of interaction model used, and how it influences the dose-toxicity surface. Furthermore, if (pre-)clinical data on drug-drug interactions are available (for instance, through drug metabolism), these can be incorporated in the interaction prior, for instance  by defining a prior for the odds multiplier at a specific dose. Note that one could also try applying more mechanistic drug interaction models with more parameters as summarized in a recent review\cite{Meyer2020} (especially if strong evidence for a particular mechanism is available), however, in the context of general Phase I Oncology dose escalation trials, the identifiability of such models would be questionable given the sparse data typically available.

According to the analytical criteria in section \ref{sec:methodology}, the presented drug-drug interaction model for the BLRM is an improvement over the previous state-of-the art linear model in logit space, as well as earlier models\citep{Thall2003}. It prevents the probability of toxicity to go to 0 when the dose of a modelled drug goes to infinity, but still supports non-monotonic combination dose-responses in scenarios with antagonistic toxicity. It could also enable running joint models for platform trials where arms with $N - 1$ drugs and $N$ drugs run simultaneously without jeopardizing accurate estimation of the marginals for the individual drug components if a suitable prior is chosen. This is now the default drug-drug interaction model in the OncoBayes2 R package from version 0.8-0 onwards.

Last, but not least, our hope is that the improved interaction model will facilitate the use of joint models, which could also make more efficient use of the data since the model structure is more appropriate for the problem at hand. The previously uncovered flaws in older interaction models prevented their use in joint models in practice, or led to them \emph{de-facto} being disabled through use of tight priors on the interaction strength parameters $\eta$ in practice. Given the improved interaction model, one can now be more sensitive to potential drug-drug interactions without risking undesired behavior at higher doses.


\section*{Acknowledgments}
We would like to thank Beat Neuenschwander and Matt Whiley for feedback on the proposed model and structure of the manuscript.

\subsection*{Author contributions}
All authors wrote and reviewed the manuscript. Lukas A. Widmer designed and performed the research. The OncoBayes2 software was co-authored by Sebastian Weber, Lukas A. Widmer and Andrew Bean, and is maintained by Sebastian Weber.

\subsection*{Financial disclosure and conflict of interest}

Lukas Widmer and Sebastian Weber are employed by Novartis Pharma AG and own stocks in the company. Andrew Bean and David Ohlssen are employed by Novartis Pharmaceuticals Corporation and own stocks in the company.

\section*{Supporting information}

The following supporting information is available as part of the online article:
\\

\noindent
\textbf{Git repository} at \url{https://github.com/luwidmer/ddi-manuscript}, containing the source LaTeX code for this manuscript as well as the R source code that fits the various models, produces the figures and compiles the manuscript.

\appendix

\nocite{*}
\bibliography{manuscript}%

\clearpage
%
%

\end{document}